\documentstyle[floats,aps,aipbook]{revtex}

\begin{document}

\title{Testing the Dipole and Quadrupole Moments of Galactic Models}

\author{Michael S. Briggs$^1$, William S. Paciesas$^1$,
Geoffrey N. Pendleton$^1$, Charles A. Meegan$^2$, Gerald J. Fishman$^2$,
John M. Horack$^2$, Chryssa Kouveliotou$^3$, Dieter H. Hartmann$^4$,
Jon Hakkila$^5$}

\address{$^1$Department of Physics \\
        University of Alabama in Huntsville, Huntsville, AL 35899 \\
     $^2$Space Sciences Laboratory \\
         NASA/Marshall Space Flight Center, Huntsville, AL 35812 \\
     $^3$Universities Space Research Association, \\
     NASA/Marshall Space Flight Center, Huntsville, AL 35812 \\
     $^4$Department of Physics and Astronomy \\
         Clemson University, Clemson, SC 29634 \\
     $^5$Department of Physics and Astronomy \\
         Mankato State University, Mankato, MN 56002}

\maketitle

\begin{abstract}
\end{abstract}

 
\section*{Introduction}

If gamma-ray bursts originate from a galactic source population, then 
at some level a galactic pattern must exist in their locations.   
Expected patterns for galactic sources are a
concentration towards the
galactic center, measured by the mean dipole moment of the locations towards the
center, $\langle \cos \theta \rangle$, where the $\theta_i$
are the angles between
the burst locations and the galactic center, or a concentration towards
the galactic plane, measured by the mean quadrupole moment about the plane,
$\langle \sin^2 b - \frac{1}{3} \rangle$, where the $b_i$ are the galactic
latitudes of the locations \cite{Pac90,Bri93}.  
To date, neither pattern has been found in the
BATSE data: the values $\langle \cos \theta \rangle$ $=0.011 \pm 0.017$
and $\langle \sin^2 b - \frac{1}{3} \rangle$ $=0.002 \pm 0.009$ (these 
values have been corrected
for BATSE's nonuniform sky exposure) for the 1122 bursts of the 3B catalog
are both consistent with zero and thus with isotropy \cite{Mee96A}.
The dominant uncertainty in these values is due to the finite sample size
\cite{Bri96}.    What galactic signatures could be hidden under these
uncertainties?

The tight limits on the quadrupole moment, in conjunction with the 
fall-off in the number of faint sources, rule out a disk origin for the majority
of the sources \cite{Mee92,Hak94,Bri95,Bri96}.    
Galactic models which remain under discussion either consist of a extended halo
or of multiple components.
A halo consistent with the data must be much larger than the solar
galactocentric distance of $R_\circ =8.5$~kpc so that the dipole moment
will be sufficiently small.   We can determine the typical scale by considering
a very simple halo: a galactocentric shell of radius $R_{\rm shell}$.
Such a shell has a dipole moment \cite{Har94}:
\begin{equation}
\langle \cos \theta \rangle = \frac{2}{3} \frac{R_\circ}{R_{\rm shell}}.
\end{equation}
Using the dipole moment of the 3B catalog (above), we obtain a $2\sigma$
lower-limit for $R_{\rm shell}$ of 120 kpc.   Any GRBs inside of this radius
will have to be balanced with sources located farther away.

In the remainder of this paper we compare galactic models which have
published moments with the observed moments of the 3B catalog.
The procedures and the models are discussed in greater detail in
an earlier work, which used a smaller sample of GRBs \cite{Bri96}.

\section*{Galactic Models}

The models with quantitative moments that we are aware of appear in Table~1.
For each model we list in the table all moments meaningfully different from
zero.   Each model listed has one or two such moments.
In some cases the parameters of the published model are based on fits
to the then existing GRB sample; conversely the parameters of some models are 
not based upon fits but are (presumably very good) examples of the model.
We have merely extracted the model moments from the publications--we have
made no effort to reoptimize the models.    Since in most cases the models
have free parameters and were created when the BATSE GRB sample was smaller,
rechoosing the parameters might improve agreement with the data.
In some cases, a full refitting might worsen agreement because of the
tightened constraints on the brightness distribution, logN-logP.

\begin{table}
\caption{Moments of Galactic Model Compared with the Observations}
\begin{tabular}{lcdd}
\multicolumn{1}{c}{Model}        &   
\multicolumn{1}{c}{Statistic}    &
\multicolumn{1}{c}{Prediction}  
\tablenote{Not corrected for BATSE's nonuniform sky exposure.}    &
\multicolumn{1}{c}{Dev.} 
\tablenote{Deviation, 
in $\sigma$, of the prediction from the value observed for the 1122
bursts of the 3B catalog (expect 109 bursts for Li et al. \cite{Li94}).  
Includes correction for sky exposure.}       \\
\tableline
\vspace{2 mm}   Eichler \& Silk \protect\cite{Eic92} \dotfill    &   
$\langle \cos \theta \rangle$     &  
0.05  &    
2.3   \\
\vspace{2 mm}   Hartmann \protect\cite{Har92} \dotfill  &
$\langle \sin^2 b - \frac{1}{3} \rangle$       &  
$-$0.05   &  
5.8    \\
\vspace{2 mm}   Li \& Dermer \protect\cite{Li92} \dotfill  &
$\langle \cos \theta \rangle$                  &   
0.048    & 
2.2    \\
\vspace{0.1 mm} Lingenfelter \& Higdon \protect\cite{Lin92}  \dotfill    &  
$\langle \cos \theta \rangle$       &
0.08    &   
4.0    \\ 
\vspace{2 mm}   &  
$\langle \sin^2 b - \frac{1}{3} \rangle$       & 
$-$0.06   & 
6.9    \\
\vspace{2 mm} Fabian \& Podsiadlowski \protect\cite{Fab93}  \dotfill        &
$\langle \cos \theta_{\rm LMC} \rangle$ 
\tablenote{Statistic is the dipole moment
to the Large Magellanic Cloud; the observed value is $-$0.010 and the predicted
sky exposure bias is $-$0.024.}     &
0.038    &  
1.4    \\
\vspace{2 mm} Smith \& Lamb \protect\cite{Smi93}:
Disk/Gaussian Shell Halo \dotfill  &
$\langle \sin^2 b - \frac{1}{3} \rangle$       &
$-$0.027   &  
3.2    \\
\vspace{2 mm} Smith \& Lamb \protect\cite{Smi93}: Dark Matter Halo/Disk\dotfill &
$\langle \cos \theta \rangle$         & 
0.057     &
2.7    \\
\vspace{2 mm} Higdon \& Ling. \protect\cite{Hig94}:
   $R_{\rm core}=7.5$ kpc, 25\% disk &
$\langle \cos \theta \rangle$      &
0.088    &   
4.5    \\
\vspace{2 mm} Higdon \& Ling. \protect\cite{Hig94}:
   $R_{\rm core}=15$ kpc, 20\% disk &
$\langle \cos \theta \rangle$            &
0.073    &
3.6    \\
\vspace{2 mm} Higdon \& Ling. \protect\cite{Hig94}:
    $R_{\rm core}=30$ kpc, 8\% disk  &
$\langle \cos \theta \rangle$          & 
0.060    &
2.9    \\
\vspace{0.1 mm} Li, Duncan \& Thompson \protect\cite{Li94} 
\tablenote{The predictions are for
bursts with 1024 ms peak flux $>$ 3.45 $\gamma$ s$^{-1}$ cm$^{-2}$, of which
there are 109 in the 3B catalog.}
\dotfill &   
$\langle \sin^2 b - \frac{1}{3} \rangle$       & 
$-$0.084   &   
1.8   \\ 
\vspace{2 mm}   &  
$\langle \cos^2 \theta - \frac{1}{3} \rangle$ 
\tablenote{The observed value of this 
quadrupole moment is $-$0.005 and the sky exposure predicted value is
$-$0.004.}      &    
0.073  &
2.6  \\
\vspace{0.1 mm} Podsiadlowski, Rees \& Ruderman \protect\cite{Pod95}:
Fig. 5a  \dotfill &   
$\langle \cos \theta \rangle$                  &
0.043    &   
1.9    \\ 
\vspace{2 mm}   &
$\langle \sin^2 b - \frac{1}{3} \rangle$       &   
$-$0.019   &
2.3   \\
\vspace{0.1 mm} Podsiadlowski, Rees \& Ruderman \protect\cite{Pod95}:
Fig. 5b  \dotfill &
$\langle \cos \theta \rangle$                  &
0.054    &
2.5    \\ 
\vspace{2 mm}  &
$\langle \sin^2 b - \frac{1}{3} \rangle$       &   
$-$0.024   &
2.9   \\
\vspace{0.1 mm} Smith \protect\cite{Smi95} \dotfill    &
$\langle \cos \theta \rangle$      &    
0.050    &   
2.3    \\
\vspace{2 mm}  &
$\langle \sin^2 b - \frac{1}{3} \rangle$       &   
$-$0.023   &
2.8   \\
\vspace{0.1 mm} Bulik \& Lamb \cite{Bul96A} \dotfill  &
$\langle \cos \theta \rangle$      &    
0.016  &
0.3 \\
\end{tabular}
\end{table}

From the table it is apparent that while some galactic models are quite
inconsistent with the observed moments, others agree well.
The best (and most recent) model \cite{Bul96A} 
is within $0.3\sigma$
of the observed dipole moment.    The model \cite{Lin92}
most distant from the data
deviates by $4.0\sigma$ from the observed dipole moment and
$6.9\sigma$ from the observed quadrupole moment.
There is an approximate trend for the moments of the more recent models
to be smaller, as additional data from BATSE has indicated that the moments
of the first post-BATSE models were too large.

Of the models including a disk component \cite{Lin92,Smi93,Hig94,Smi95}, the
one that best matches the data is the Dark Matter Halo/Disk model of
Smith \& Lamb \cite{Smi93}, which has a $2.7\sigma$ deviation in
$\langle \cos \theta \rangle$.    The largest moment in this model
is the dipole moment of its halo component, since only 20\% of the bursts
originate from the disk component.

The remaining models all assume that GRBs originate from an extended
halo.  These models fall into
two classes: arbitrary models which postulate a source radial distribution
and high-velocity neutron star (HVNS) models which assume that the halo
consists of HVNS ejected from the disk.
The HVNS models have the advantage of being based upon plausible sources
and incorporating more physics, but have potential difficulties
explaining why only HVNS burst and whether there are sufficient HVNS
to produce the observed burst rate.

The models \cite{Eic92,Har92,Hig94}
which postulate a source distribution usually
assume a dark matter halo form, following the example of Paczy\'{n}ski
\cite{Pac91}.
To match the data, these models are driven to very large core radii, 
larger than assumed in dark matter models of the galactic rotation curve
and larger than any observed galactic component \cite{Bri96}.
Also suggested are a Gaussian shell halo
\cite{Smi93} and a exponential halo \cite{Smi93,Smi95},
both of which differ from any known galactic population.

The first HVNS model \cite{Li92} is still in acceptable agreement with
the data, probably because of the 1000 km~s$^{-1}$ velocity assumed for
all of the bursting sources, a higher value than used by more recent
versions of this model.    The most recent HVNS model \cite{Bul96A}
closely matches the data.    The unusual model of Fabian and Podsiadlowski
\cite{Fab93} is also in good agreement with the data despite using
an unusally low source velocity, 400 km~s$^{-1}$.   This is achieved
by the unique assumption that GRB sources are born only in the Magellanic
Clouds, so that the sources are born at halo distances and 
can easily escape their birth site.   However, if sources are born
in the disk of the Milky Way
at even a small fraction of the Magellanic Cloud birth rate
per mass, a strong disk signature would be produced.

Since the uncertainties on the observed 
moments decrease as $1/\sqrt{N_B}$, where
$N_B$ is the number of burst locations in the sample, further progress
in testing the moments of galactic models will be slow.
Collecting additional data still has several important benefits: it will
tighten the constraints on galactic models and aid in analyzing
suggested sub-classes of bursts \cite{Bel95,Kou96,Pen96}. 
The tightened limits on
the properties of a hypothetical Milky Way halo will aid in interpreting
proposed observations of the corresponding halo of 
 M31, observations which are intended to distinguish between 
halo and cosmological distance scales \cite{Bul96B,Har96A,Har96B,Li96,Mee96B}.

\end{document}